\documentclass[aps,pra,preprint]{revtex4}

\usepackage{graphicx}
\usepackage{dcolumn}
\usepackage{bm}

\usepackage[unicode]{hyperref}
\usepackage{upgreek}
\usepackage[dvipsnames]{xcolor}
\usepackage{amsmath}
\usepackage[utf8x]{inputenc}
\usepackage{braket}
\usepackage[english]{babel}
\usepackage{enumitem}
\usepackage{natbib}
\usepackage[normalem]{ulem}

\begin{document}

\title{Efficient loading of thulium atoms in a compact MOT for a transportable optical clock}

\author{A.\,Golovizin}       
\email{artem.golovizin@gmail.com}

\author{D.\,Tregubov}

\author{D.\,Mishin}

\author{D.\,Provorchenko}
\affiliation{P.N.\,Lebedev Physical Institute, Leninsky prospekt 53, 119991 Moscow, Russia}

\author{N.\,Kolachevsky}
\affiliation{P.N.\,Lebedev Physical Institute, Leninsky prospekt 53, 119991 Moscow, Russia}
\affiliation{Russian Quantum Center, Bolshoy Bulvar 30,\,bld.\,1, Skolkovo IC, 121205 Moscow, Russia}
\date{\today}

\begin{abstract}

We report on building of a compact vacuum chamber for spectroscopy of ultracold thulium and trapping of up to 13 million atoms.
Compactness is achieved by obviating a classical Zeeman slower section and placing an atomic oven close to a magneto-optical trap (MOT), specifically at the distance of 11 cm. 
In this configuration, we significantly gained in solid angle of an atomic beam, which is affected by MOT laser beams, and reached 1 million atoms loaded directly in the MOT.
By exploiting Zeeman-like deceleration of atoms with an additional laser beam, we increased the number of trapped atoms to 6 million.
Then we gained an extra factor of 2 by tailoring the MOT magnetic field gradient with an additional small magnetic coil.
Demonstrated results show great perspective of the developed setup for realizing a compact high-performance optical atomic clock based on thulium atoms.
\end{abstract}

\maketitle

\section{Introduction}
\label{Section:intro}

Rapid progress in the field of high-precision optical atomic clocks translated into numerous applications like relativistic geodesy and gravimetry \cite{Grotti2018,Takamoto2020test}, global timekeeping and synchronization \cite{Riehle2017,Yao2019Optical-Clock-BasedScale}, navigation \cite{Giorgi2019AdvancedGeodesy} as well as delicate tests of fundamental physics \cite{Kennedy2020precision,Sanner2019optical,Wciso2018first,BACON2021Frequency}. 
To satisfy the growing demands \cite{Riehle2018},  transportable systems with relative systematic uncertainty and instability at the $10^{-17}$ level or better are required. 
Besides technical difficulties, one of the main challenges is to establish tight control over the environmental influence on the clock transition frequency. 
The leading perturbation factors are the  external magnetic  and electric fields \cite{Poli2014,Ludlow2015} (including blackbody radiation (BBR)), which can shift the frequency at the $10^{-14}$ level and require a special approach to compensate for.

Single-ion-based optical clocks are typically more robust  compared to atomic clocks based on neutral atoms because of the deep trapping potential, long ion life time and small BBR shift for the ion systems.  
The ultimate relative instability and inaccuracy for both clock types is now approaching the $10^{-18}$ level \cite{Huntemann2016, Nicholson2015systematic,McGrew2018atomic}, however, single-ion clocks require much longer averaging time due to lower statistics. 
Indeed, neutral-atom clocks  typically operate with  ensembles of $10^3$-$10^4$ neutral atoms in an optical lattice, compared to a single ion in a Paul trap.
For many applications like geodesy and navigation, averaging time plays a crucial role which points towards neutral atoms being a favorable platform.
On the other hand, neutral atom clocks typically possess higher sensitivity to external electromagnetic (EM) fields compared to ion clocks.
Another challenge is to design a compact and robust vacuum system.
An intensive flux of initially hot atoms should be decelerated and loaded into a magneto-optical trap (MOT), which is commonly done with the help of a Zeeman slower \cite{Phillips1982Zeeman}.  
Moreover, a delicate optical cavity forming an optical lattice  should be incorporated  into the vacuum system. The transportable Sr optical clocks \cite{Koller2017,Takamoto2020test} possess a few-m$^3$-size which is still significantly more compared to the demonstrated ion-based systems \cite{Cao2016,Zalivaco2020compact}. 
To achieve the ultimate clock performance, cryogenic vacuum techniques are used to provide suppression of the  BBR shift \cite{Ushijima2015,Takamoto2020test}.
Therefore, it would be desirable  to combine the  advantages of ion-based clocks (small size, robustness and low sensitivity to EM fields) with those of neutral atom clocks (good statistics, low short-time frequency instability).

We have experimentally demonstrated a number of attractive features of  the inner-shell magnetic-dipole transition in neutral thulium at 1.14\,$\mu$m \cite{sukachev2016inner,Golovizin2019inner}.  
Its frequency possesses  very low sensitivity to BBR, namely 3000 times lower compared to the Sr metrological transition at 698 nm. The magic  wavelength is close to 1064\,nm where powerful low-phase-noise lasers are available. 
Moreover, the bicolor operation \cite{Fedorova2020Simultaneous, Golovizin2021Bicolor}  based on simultaneous interrogation of two hyperfine transitions allows us to  cancel out the quadratic Zeeman shift. 
Our analysis \cite{golovizin2020estimation} shows  that  Tm optical lattice clocks at room temperature can reach the low $10^{-18}$ level of total systematic frequency shift. 
This drastically softens requirements for the external parameters like thermal surroundings, magnetic field stability, as well as the spectral purity and frequency stability of the optical lattice laser. 
All these features are highly desirable for development of a transportable optical clock operating below the $10^{-17}$  level of fractional instability and inaccuracy. 
In this paper, we focus on development of a compact and robust thulium MOT assembly which is the key part of a clock system.

For alkali atoms with high vapor pressure (like Rb), compact MOTs with high loading rates  have been  known for decades, and great progress in their miniaturization was achieved recently with the pyramid \cite{arlt1998pyramidal, bodart2010cold} and grating \cite{nshii2013surface, barker2019single} configurations.
However, most of the atomic species used today for the high-performance optical clocks have low vapor pressure at room temperature, so  high-temperature ovens are necessary. 
The oven is typically placed far apart from the MOT region for two reasons. 
First, a Zeeman slower of a certain length (a few tens on cm) is typically required  to decelerate atoms from the thermal speed to the MOT's capture velocity. Second, thermal radiation from the hot oven induces an additional BBR shift of the clock transition frequency (proportional to $T^4$, $T$ is the temperature). Proximity of the oven also induces undesirable collisions.

In recent years, significant progress has been made on the route of compactization of the optical clock setups.
To begin with, Zeeman-free systems with a  2D-MOT for initial slowing of atoms \cite{Barbiero2020sideband,Nosske2017twodimensional} were demonstrated. 
These configurations significantly reduce the system size and lift up some undesirable effects (like hot atomic beam, direct view of the oven, power consumption of Zeeman coils or magnetic field from permanent magnets), but they still require an extra volume for pre-cooling and increase complexity of the system.
Beside this, the pyramid \cite{Bowden2019pyramid} and grating \cite{Sitaram2020grating} MOTs for Sr atoms were developed.
An interesting approach with laser ablation of atoms was implemented in refs.\,\cite{Yasuda2017lasercontrolled, Kock2016laser}.

In this work, we demonstrate experimental realization of  a compact thulium MOT without the use of the ``classical'' Zeeman slower section. The MOT is  efficiently loaded directly from the atomic oven placed in the proximity of the MOT region. 
Very low sensitivity of the 1.14\,$\mu$m clock transition frequency to BBR ensures small impact of the oven radiation. 
We simulated the atom capture process in our setup in order to optimize the magnetic field configuration and to evaluate the MOT's loading rate. Two different configurations with  6 or 7 laser beams (one additional beam for deceleration) where analyzed, both offering prospects of the loading rate of more than $10^{6}$\,atoms/s. 

The manuscript is organized as follows: the Section\,\ref{Section:setup} describes the vacuum chamber geometry and laser cooling beams configuration. 
The Section\,\ref{Section:modeling} is devoted to a brief description of our numerical model.
In the Section\,\ref{Section:results}, we demonstrate experimental  results.
Finally, we sum up the conclusions in the Section\,\ref{Section:conclusion}.

\begin{figure*}
\center{
\resizebox{0.75\textwidth}{!}{
\includegraphics{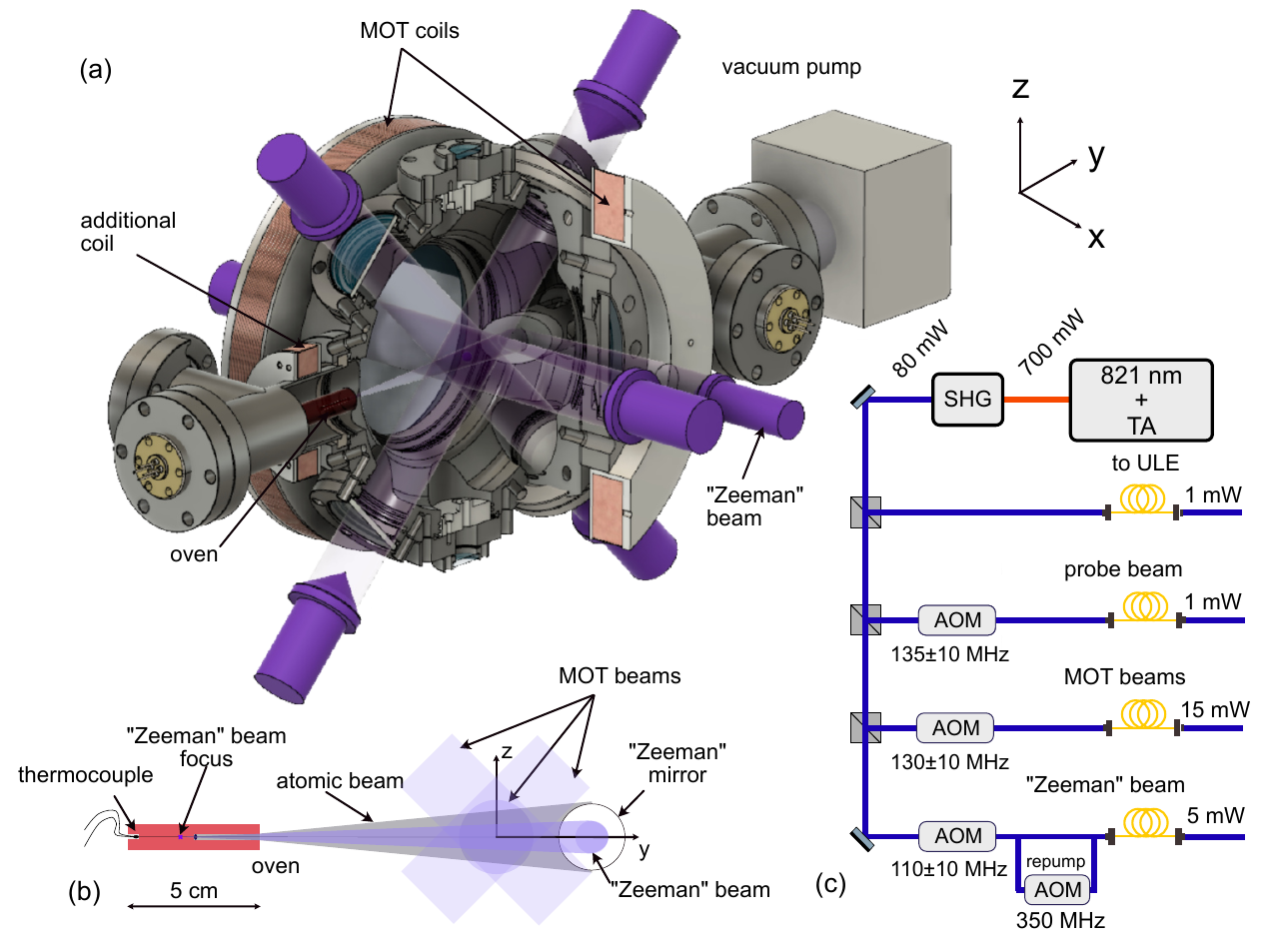}
}
\caption{
The experimental setup. (a) The  3D model of the experimental vacuum chamber. 
The cooling  (MOT) and decelerating (``Zeeman'') laser beams are shown with purple arrows.
(b) The sketch of the atomic oven and the cooling beams.
(c) The optical scheme. The output of the second harmonic generation (SHG) cavity at 410.7\,nm is split in four beams. 
The MOT, ``Zeeman'' and probe laser beams are frequency shifted  by  the acousto-optic modulators (AOMs) and  delivered to the vacuum chamber with the optical fibers (yellow curves).
The repumper radiation $\ket{g,F=3}\rightarrow\ket{b,F=4}$ is added to the ``Zeeman'' beam.}
\label{fig:scheme}}
\end{figure*}

\section{Setup}
\label{Section:setup}

The main part of the the experimental apparatus is shown in Fig.\,\ref{fig:scheme}.a.
We use a 6-inch spherical octagon vacuum chamber from Kimball Physics Inc.\,\cite{kimball} . 
The atomic oven is mounted inside the CF40 T-flange.
The non-evaporable getter pump Z100 from SAES Group\,\cite{saes}, the angle valve and the electrical feedthrough for the in-vacuum-mounted optical lattice enhancement cavity are connected to the main chamber with the  CF40 cross. 
One of the 3 orthogonal pairs of the MOT beams goes through the center of CF100 windows along the $x$ axis, while the other two orthogonal beam pairs lay in the $z-y$ plane and pass through CF40 windows.
This configuration is further referred to as the 6-beam MOT.
In the 7-beam MOT, we use an additional, focused ``Zeeman'' laser beam which
is parallel to the $x$ axis. It is deflected by an in-vacuum elliptical-shape mirror opposite to the atomic beam emitted by the oven. This allows us to achieve  good ``Zeeman'' beam convergence on the oven and prevent optical windows from metal deposition.
The probe laser beam goes in the horizontal plane and is not shown. 
Coils for the MOT's magnetic quadrupole field are mounted on the CF100 flanges, each coil containing 546 loops of a 1-mm diameter copper wire.
The magnetic field gradient at the MOT center equals 14\,G/cm (along the $x$ axis) per 1\,A  current. 
An additional pair of magnetic coils is installed axially to the $y$ direction which allows one to modify the magnetic field profile along the atomic beam, which is  discussed in  the Sec.\,\ref{Section:7beam}.

A more detailed sketch of the atomic oven is shown in Fig.\,\ref{fig:scheme}.b.
The oven is made from a 50\,mm-long, 10\,mm-diameter ruby rod. 
Metallic thulium is loaded into a $L=25$\,mm-deep hole with a diameter of $d=2.5$\,mm.
The atomic beam divergence is approximately $\Delta\theta = d/L = 0.1$\,rad. The oven surface is homogeneously heated by a  $0.2$\,mm-diameter tungsten wire. 
A thermocouple is attached to the ruby rod to measure the oven temperature.  
To achieve an atomic flux of  $\Phi\approx10^{10}$\,atoms/s at the oven's exit,  one needs to heat the oven to  $500\,^{\circ}$C, which requires 20\,W of electrical power.

The optical part of the setup is shown in Fig.\,\ref{fig:scheme}.c. 
For laser cooling of thulium atoms we use a strong cooling transition $\ket{g,F=4}\rightarrow\ket{b,F=5}$ at 410.7\,nm with a natural linewidth of $\gamma=10$\,MHz \cite{Sukachev2010}.
Here, $\ket{g}$ denotes the ground level and $\ket{b}$ --- the upper cooling level, $F$ is the total momentum of the atom.
A 821\,nm semiconductor laser with a tapered amplifier (Sacher Lasertechnik Group \cite{sacher}) and a home-made second harmonic generation cavity \cite{shpakovsky2016compact} provides 80\,mW of 410.7\,nm radiation. 
The laser frequency is stabilized to a ULE cavity mode.
The probe, ``Zeeman'', repumping and MOT cooling beams are frequency shifted and modulated by corresponding acousto-optic modulators (AOMs).
Optical radiation is delivered to the vacuum apparatus by 3 optical fibers. 
In this configuration, we get 15\,mW radiation at 410.7 nm for the MOT beams and 1\,mW for the resonant probe beam. 
The MOT beams radii ($1/e^2$ intensity level) equal $w_\textrm{MOT}=14$\,mm. 
The input beams apertures are limited by $30$\,mm collimation lenses, while the back-reflected beams are truncated by $22$\,mm-diameter quarter-waveplates.
The probe beam has a radius of 1.15\,mm.
Frequency of the probe beam was set close to the exact transition resonance.

\section{MOT loading simulations}
\label{Section:modeling}

We performed Monte-Carlo trajectory simulation of atoms   defused from the atomic oven interacting with the cooling laser beams. 
Our approach is similar to the one described in Ref.\,\cite{Barbiero2020sideband}.
For every set of parameters, we calculated $n_\textrm{tot}=1000$ atomic trajectories using the 4th-order Runge-Kutta algorithm with a time step of $\delta t=50\,\mu$s. 
We assumed that an atom was trapped if its final coordinate was closer than  $0.5\,w_\textrm{MOT}$ from the MOT center, where $w_\textrm{MOT}$ is the  MOT cooling beams radii.

\subsection{Initial parameters}
The initial position of atoms was fixed at the oven center $\mathbf{r_0}=(0,y_\textrm{oven},0)$. Here,  $y_\textrm{oven}=-11.4$\,cm is the distance from the MOT center to the oven (see Fig.\,\ref{fig:scheme}.b). 
The initial velocity vector was chosen randomly from the uniform distribution within a cone with an opening angle of $2\Delta\theta$. The velocity modulus was sampled from the  Maxwell-Boltzman distribution for the oven temperature $T=500^\circ$C truncated at $v_\textrm{th}$.
The velocity threshold $v_\textrm{th}$ was estimated by analyzing the capture process of atoms moving along the  $y$ axis, i.e. with the initial velocity $\boldsymbol{v}=(0,v,0)$. 
For different MOT parameters, $v_\textrm{th}$ lay between 35 and 150 m/s.
The capture efficiency was inferred as $\zeta = n_\textrm{trapped}/n_\textrm{tot} \times \eta(v_\textrm{th})$, where $\eta(v_\textrm{th})$ is the fraction of atoms in the truncated part of the Maxwell-Boltzman distribution.

The total atomic flux was evaluated using an expression for a capillary output:
\begin{equation}\label{capillary}
    \Phi = \frac{4\sqrt{\pi}}{3}\frac{n_\textrm{ov}v_\textrm{th}d^3}{8L},
\end{equation}
where $n_\textrm{ov}$ is the thulium atomic vapor density \cite{SukachevDisser} and $v_\textrm{th}$ is the most probable thermal velocity at $T=500^\circ$C.
This assumption may differ from our experimental situation because the expression (\ref{capillary}) corresponds to the case when the capillary of length $L$ and diameter $d$ is connected to a reservoir of atomic vapor at a certain temperature. 
In our case, a chunk of sublimating metal is placed in the capillary. 

\subsection{Atom-light interaction}
Atom-light interaction was considered as a classical friction-type force induced by the photon scattering on a 2-level-system:
\begin{equation}
    \boldsymbol{F} = \hbar k \sum_i R^\textrm{sc}_i \boldsymbol{\hat{k}}_i + \hbar k \sqrt{N^\textrm{sc}}\boldsymbol{e_\textrm{r}}
    \label{eq:force}
\end{equation}
Here, $\hbar$ is the reduced Plank constant, $k=2\pi/\lambda_{410}$. Summation goes over all optical beams (see Fig.\,\ref{fig:scheme}) with the unit wave vectors $\boldsymbol{\hat{k}}_i = \boldsymbol{k}_i/k$ and the $i$th beam scattering rate of $R^\textrm{sc}_i$. 
The last term is responsible for heating from spontaneous emission. It introduces a momentum kick in a random direction $\boldsymbol{e_\textrm{r}}$ proportional to $N^\textrm{sc} = \delta t \sum_i R^\textrm{sc}_i$.

For every time stamp we calculated $i$th beam scattering rate
\begin{equation}
    R^\textrm{sc}_i = \frac{\Gamma}{2} \frac{s_i}{1 + s_i + 4(\Delta_i/\Gamma)^2}
\end{equation}
based on the atom's coordinate $\boldsymbol{r}$ and velocity $\boldsymbol{v}$ from the previous step.
Here, $s_i = s_i^0 \exp\left(-\frac{2\left|\boldsymbol{r}\times\boldsymbol{\hat{k}}_i\right|^2}{w_i^2}\right)$ is the saturation parameter for the $i$th beam with a waist of $w_i$ and on-axis saturation parameter of $s_i^0$.
For the MOT beams, $w_i$ and $s_i^0$ are constant, while for the ``Zeeman'' beam they depend on the $y$ coordinate.
In order to take into account finite beams apertures, we set $s_i$ to zero when $\left|\boldsymbol{r}\times\boldsymbol{\hat{k}}_i\right| \ge r_\textrm{lens}$. Effective detuning of the $i$th beam from the exact resonance is calculated as $\Delta_i = \Delta_i^0 + (\boldsymbol{k}_i, \boldsymbol{v}) - g_b\mu_B/\hbar\, B(\boldsymbol{r}) \sigma_i$.
Here, $\Delta_i^0$ is the frequency detuning of the $i$th beam (either $2\pi\Delta\nu_\textrm{MOT}$ or $2\pi\Delta\nu_\textrm{Z}$), $g_b=1$ is the Lande g-factor of the upper cooling level, $B(\boldsymbol{r})$ is the magnetic field and $\sigma_i = \pm 1$ is the coefficient related to the beam polarization.
One can not consider cooling beams independently when  $R^\textrm{sc} = \sum_i R^\textrm{sc}_i$ is larger than $\Gamma/2$ (the maximum photon scattering rate on the transition). 
To avoid this, we used normalized scattering rates $\Tilde{R}^\textrm{sc}_i = R^\textrm{sc}_i / (1 + 2(R^\textrm{sc} - R^\textrm{sc}_i)/\Gamma) $ in Eq.\,\ref{eq:force}.

\section{Results}
\label{Section:results}

\subsection{6-beam MOT}
\label{Section:6beam}

Simulations predicted that about $10^{-5}$ fraction of total atomic flux from the oven could be trapped in the 6-beam MOT.
This corresponds to the  loading rate of $10^5 - 10^6$\,atoms/s.
The pulse sequence of the experiment is the following: the MOT is loaded during $t_\textrm{load}=0.2-4$\,s with the MOT laser beams and quadrupole magnetic field on. 
We switch off the MOT laser beams (using AOM) and magnetic field, and  after 1\,ms interval the  number of atoms is measured  using  fluorescence signal from the resonant 410\,nm beam  recorded by a CMOS camera.
Then we wait for $t_\textrm{wait}=0.5-1$\,s before the next loading cycle.

\begin{figure}
\center{
\resizebox{0.8\textwidth}{!}{
\includegraphics{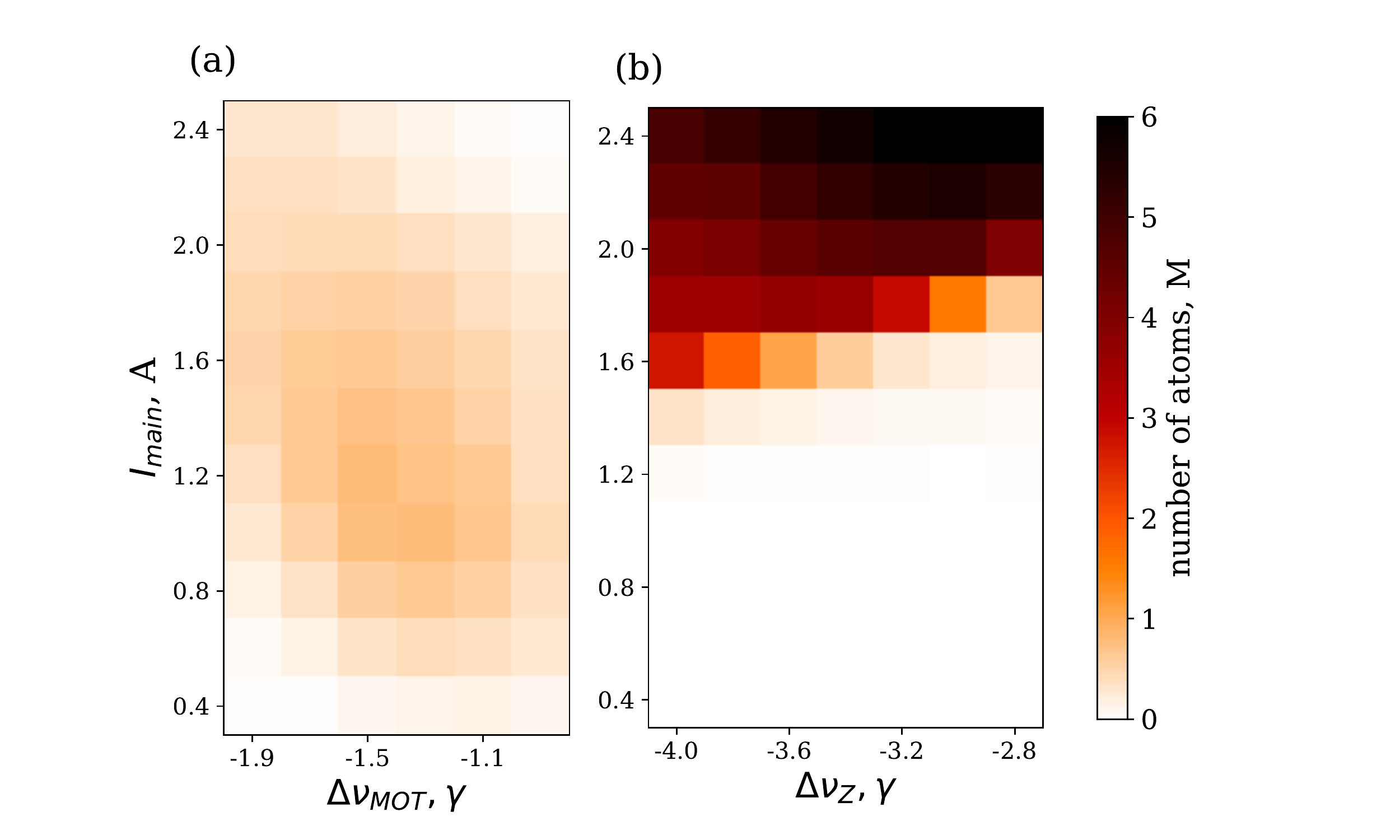}
}
\caption{
(a) Experimentally measured number of atoms in the 6-beam MOT as a function of the current flowing through the MOT magnetic coils (proportional to the field gradient)  and  the frequency detuning of the MOT cooling beams $\Delta\nu_\textrm{MOT}$.  
(b) Similar plot for the  7-beam MOT, but in this case we scan the frequency detuning of the Zeeman beam $\Delta\nu_\textrm{Z}$. The color bar  is common for both plots. 
}
\label{fig:motNatoms}}
\end{figure}

We measure the  number of atoms in the 6-beams MOT after $t_\textrm{load}=3$\,s loading time as a function of the cooling beams frequency detuning and value of the quadrupole magnetic field, as it is shown in Fig.\,\ref{fig:motNatoms}.a. 
The  frequency is scanned using a single-pass AOM.
To compensate for variations of the diffraction efficiency and optical fiber coupling, we apply feed-forward on the AOM's RF drive power according to the calibration. 
In this configuration we trap up to $10^6$ Tm atoms.

\subsection{7-beam MOT}
\label{Section:7beam}

Regarding the number of atoms, the 6-beam MOT can already be used for optical clock applications. 
However, our simulations show that the MOT magnetic field gradient can be used for a Zeeman-like slowing of the atoms escaping the oven with an additional ``Zeeman'' laser beam (see Fig.\,\ref{fig:scheme}.b)  directed opposite to the atomic flux.
We use the ``Zeeman'' beam with a convergence of  $\Theta=0.1$\,rad (2 times smaller than the atomic beam divergence, currently limited by the out-of-vacuum optics).  
The  beam radius at the MOT center equals $w_\textrm{Z} = 5$\,mm.
The implemented configuration allows one to:
\begin{enumerate}
    \item reduce blowing-out of atoms from the MOT by the ``Zeeman'' beam.
    \item decelerate atoms opposite to their velocities instead of along the $y$ axis (the ``Zeeman'' beam direction).
\end{enumerate}

The pulse sequence is  the same as for the 6-beam MOT.
After certain  optimization, we measure the number of trapped atoms in the 7-beam MOT for different magnetic field gradients and ``Zeeman'' beam detunings as shown in Fig.\,\ref{fig:motNatoms}.b.
We observe a relatively sharp edge which separates the region with  efficient trapping from the region where almost no atoms are captured. 
This corresponds to the  resonance condition for the ``Zeeman'' laser beam at a certain magnetic field gradient.
In this configuration we reached 6 millions atoms in the MOT.

\begin{figure}
\center{
\resizebox{0.7\textwidth}{!}{
\includegraphics{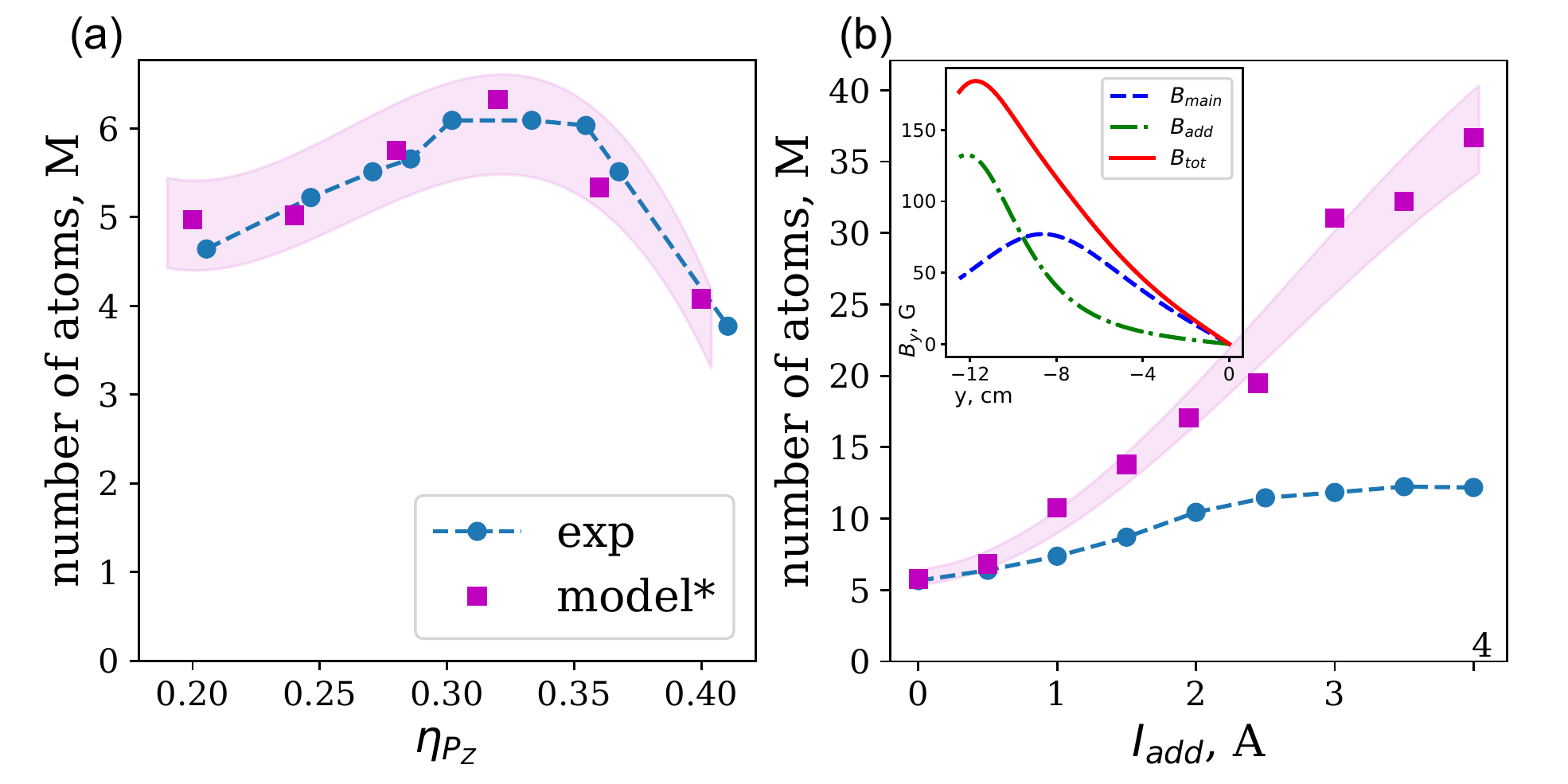}
}
\caption{
(a) The maximum  number of atoms in the 7-beam MOT as a function of fraction of optical power in the ``Zeeman'' beam $\eta_{P_\textrm{Z}}$. 
(b) The number of atoms as a function of current $I_\textrm{add}$  through the additional coil pair at $I_\textrm{main}=2.4$\,A.
The experimental data is shown in blue circles, the simulation is shown in red squares, the estimated uncertainty (statistical) of the simulation is depicted with the shaded area. 
Inset in (b) shows the magnetic field profile along the $y$ axis produced by the main coils at $I_\textrm{main}=2.4$\,A (blue dashed line), additional coils at $I_\textrm{add}=4$\,A (green dashed-dotted line), and their sum (red solid line).
}
\label{fig:Badd}}
\end{figure}

We also study dependency of the trapping efficiency on the  fraction of optical power in the ``Zeeman'' beam $\eta_{P_\textrm{Z}} = P_\textrm{Z}/(P_\textrm{Z} + P_\textrm{MOT})$ (Fig.\,\ref{fig:Badd}.a). 
Starting from $\eta_{P_\textrm{Z}}=0.35$, it  rapidly decreases  since the ``Zeeman'' beam pushes equilibrium cloud position outside intersection of the MOT beams. Simulations shown in the same figure confirm the observed behavior within  the simulation uncertainty (the shaded area).
The number of trapped atoms in all simulations are  scaled by a factor of $\xi=0.8$ to achieve the best  matching with the experimental data shown in Fig.\,\ref{fig:Badd}.a.

One can see form Fig.\,\ref{fig:motNatoms}.b that the increase of current  $I_\textrm{main}$ in the 7-beam MOT results in the continuous growth of the number of atoms (which is also confirmed by simulations).
In our case, it was limited by technical reasons since $I_\textrm{main}=2.4$\,A corresponds to  about 100\,W power dissipation which is already at the edge of the practical use.
As an alternative approach to increase $I_\textrm{main}$ further, we used the additional coil pair (see Fig.\,\ref{fig:scheme}.a). 
These coils modify the magnetic field profile along the $y$ axis making its gradient more uniform as it is shown in the inset of Fig.\ref{fig:Badd}.b.
According to our simulations, this should raise by $20-40$\,m/s the maximum trapping velocity of atoms and significantly increase the loading rate.

Working in the optimal configuration (determined from the Fig.\,\ref{fig:motNatoms}.b),  we observe a two times increase of the number of atoms in MOT when changing the additional coils current $I_\textrm{add}$ from 0 to 4\,A, as shown in Fig.\ref{fig:Badd}.b.
It is much less than expected from the model (red dots and the shaded area). 
The discrepancy is most probably due to the Zeeman structure of thulium energy levels.
This issue is considered in more details in the Supplementary Material.

\subsection{Loading dynamics}
\label{Section:dynamics}

\begin{figure}
\center{
\resizebox{0.5\textwidth}{!}{
\includegraphics{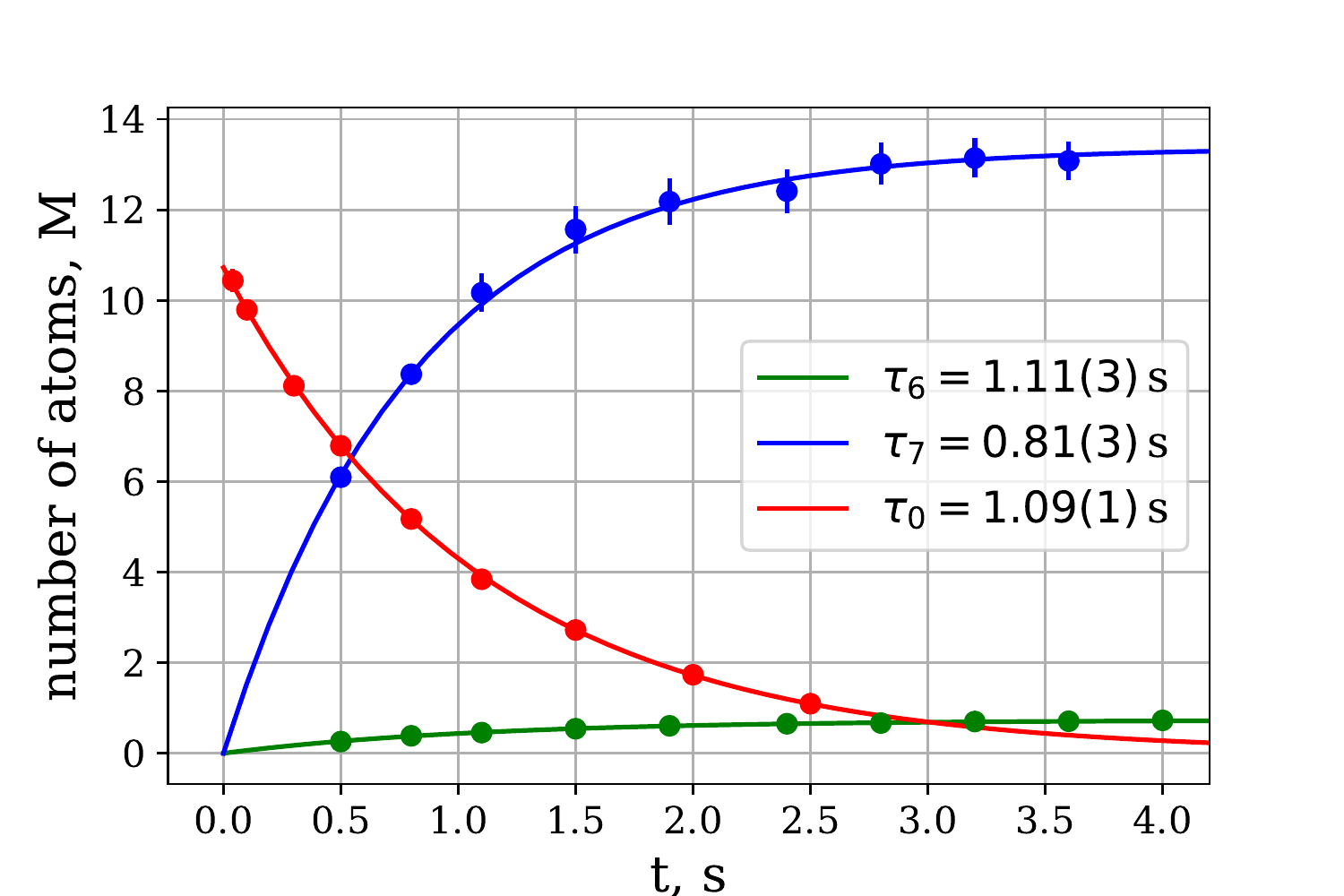}
}
\caption{
The loading and the loss dynamics of the 6-beam (green) and the 7-beam (blue and red) MOTs.
The legend shows time constants inferred from the fits (see text). 
Number of atoms is in millions.
}
\label{fig:loading}}
\end{figure}

The loading dynamics at the optimal parameters for the  6-beam and 7-beam MOTs are presented in  Fig.\,\ref{fig:loading} with green and blue dots, correspondingly. 
The experimental data are fitted by the exponential function $N_\textrm{load}(t) = R\times \tau (1-e^{-t/\tau})$ giving the time constant $\tau_6 = 1.11(4)$\,s  and loading rate $R_6=0.66(2)\times10^{6}$\,atoms/s for the 6-beam MOT, and $\tau_7 = 0.81(4)$\,s and $R_7=16.5(5)\times10^{6}$\,atoms/s for the 7-beam MOT.
The lifetime of atoms in the MOT is measured by  switching off the 7-beam MOT loading (turning off the ``Zeeman'' beam) and measuring the  number of atoms as a function of holding time (red dots).
The experimental data is fitted by the exponential decay function $N_\textrm{loss}(t) = N_0 e^{-t/\tau_\textrm{0}}$ (red solid line).  
The MOT lifetime equals $\tau_\textrm{0} = 1.09(1)$\,s, which agrees well with the loading time constant $\tau_6$. 
The observed atoms lifetime in the MOT also indicates proper vacuum conditions sufficient for interrogation of the  clock transition in the optical lattice.

\section{Conclusion}
\label{Section:conclusion}

On the way to transportable optical clock operating on the inner-shell transition at 1.14\,$\mu$m in Tm, we designed and built the compact MOT system without a traditional Zeeman slower section and demonstrated trapping of up to 13 millions atoms. 
In this setup, we can load about 1 million atoms just in 100\,ms which can considerably increase duty cycle of the compact clock setup. 
The observed lifetime of atoms in the MOT exceeds 1 second which points to proper vacuum conditions and low collision rate with atoms from the oven. To prevent an envisioned impact of thermal radiation and atomic flux from the hot oven on the clock transition frequency,  we install a  mechanical shutter between the oven and the MOT.
With all lasers being available from our ``stationary'' system, we expect rapid implementation of further steps needed to run the optical clock in the developed setup: deep second-stage laser cooling, trapping of atoms into the optical lattice and clock transition spectroscopy.

We believe that the number of trapped atoms can be significantly increased, if necessary (e.g. for experiments with Bose-Einstein condensates \cite{Davletov2020bose}),  with the further straightforward steps: raising the oven temperature, increasing the optical power per each laser beam and the convergence angle of the ``Zeeman'' beam. 
The ongoing development of deep laser cooling of thulium down to $1\,\mu$K \cite{Provorchenko2021investigation} may open new applications for cold thulium atoms, e.g. as a platform for atomic interferometry and  gravimetry.

\section{Acknowledgements}

The authors acknowledge the support of RSF Grant No. 19-72-00174. 
We are grateful to Vadim Sorokin for invaluable help with building the thulium oven.

\bibliography{references, referencesMendeley}

\section*{Supplemental Material}
\label{Section:discussion}

From our simulations we optimized the trap configuration, namely, the  beams directions, radii and the expected frequency detunings, as well as the magnetic field gradient. 
It allowed us to properly  design the vacuum chamber and the optical setup. Numerical simulations  gave us valuable hints on the ways to increase the number of trapped atoms such as implementation of the converging ``Zeeman'' beam and additional magnetic coils to optimize the  magnetic field gradient.
Taking into account simplifications used in the modeling, we expect only modest  agreement  between the theory and experiment. 

The model does not  take into account the  following aspects:
\begin{enumerate}[label=$S{\arabic*}$.]
    \item the Zeeman structure of the ground and the upper cooling levels;
    \item deviation of the laser beam polarization from  $\sigma^+$ outside the beam axis, where the magnetic field direction is not collinear with the beam wave vector;
    \item the decay channels from the upper level to the  intermediate levels;
    \item off-resonant excitation of the $\ket{b,F=4}$ state and  subsequent decay to the $\ket{g,F=3}$ level;
    \item  deviation of the atomic flux from the capillary velocity distribution.
\end{enumerate}

\begin{figure}
\center{
\resizebox{\textwidth}{!}{
\includegraphics{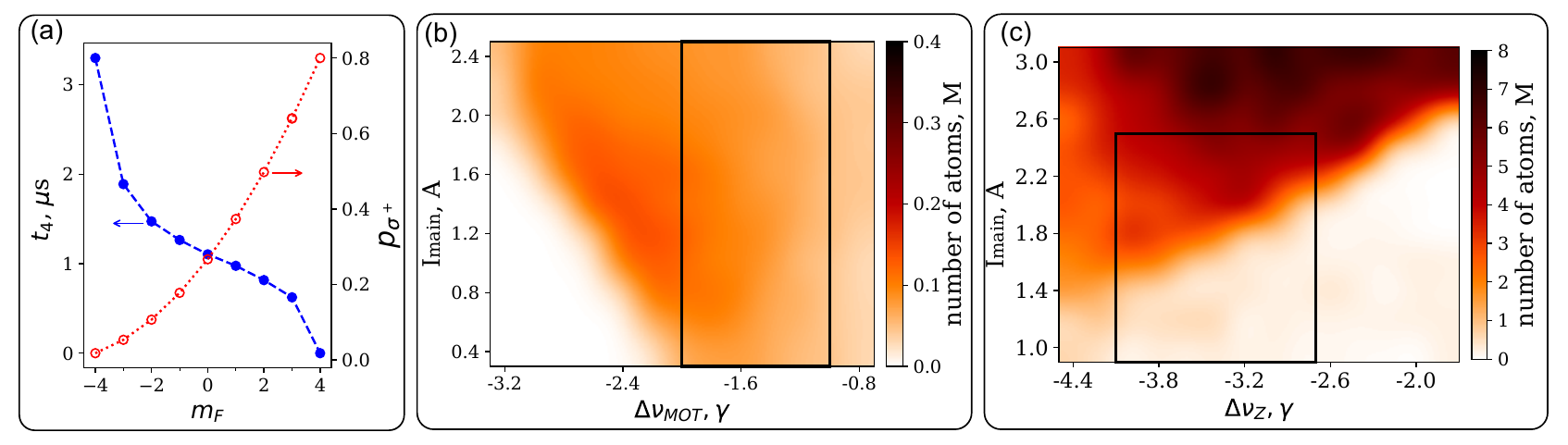}
}
\caption{
Results of the simulations. 
(a) Red empty circles: the probability $p_{\sigma^+}$  of $\sigma^+$ transitions relative to the 2-level atom case for different ground state magnetic sublevels $m_F$. 
Blue filled circles: time $t_4$ necessary for an atom to be optically pumped to $m_F=+4$ state for our experimental conditions.
(b) Prediction of the number of atoms trapped in the 6-beam MOT as a function of the frequency detuning $\Delta\nu_\textrm{MOT}$ and current through the MOT coils $I_\textrm{main}$.
Black box shows the region corresponding to experimentally covered range (Fig.\,\ref{fig:motNatoms}.a).
(c) Similar results for the  7-beam MOT. The frequency detuning $\Delta\nu_\textrm{Z}$ corresponds to the ''Zeeman'' beam.
Color bars represent the number of trapped atoms in millions.
}
\label{fig:modelData}}
\end{figure}

Note, that the simulations reasonably  well reproduce the experimentally observed number of trapped atoms as a function of the parameter $\eta_{P_\textrm{Z}}$ (Fig.\,\ref{fig:Badd}.a). 
One also observes good agreement for the ``Zeeman'' beam detuning  and magnetic field gradient dependencies (Fig.\,\ref{fig:motNatoms}.b and  Fig.\,\ref{fig:modelData}.c) (7-beam MOT configuration). 
In these cases the atom-light interaction seems to be close to that of the 2-level system and the  effects which are discussed below do not  play a significant role.

Noticeable disagreement between simulation and experiment are observed in the following cases:
\begin{enumerate}
    \item The capture efficiency {\it{vs.}} $I_\textrm{add}$ (Fig.\,\ref{fig:Badd}.b).
    It is probably associated with neglecting the Zeeman levels structure in the model.  The frequencies of  
    $\sigma^+$ (as well as $\sigma^-$ and $\pi$) transitions from different $m_F$ states are almost equal in the presence of the magnetic field owing to close values of Lande g-factors of $\ket{g,F=4}$ ($g_g = 1.00$) and $\ket{b,F=5}$ ($g_b=1.01$) levels. 
    Still, the  strength of the transitions differ significantly as it is shown in  Fig.\,\ref{fig:modelData}.a (red empty circles) and the atom-field interaction for low $m_F$ states is weaker compared to  the 2-level-atom model.
    Correspondingly, the time needed for an atom in the particular $m_F$ state to be optically pumped to the $m_F=+4$ state (with $90\%$ probability) increases for lower $m_F$ states as  shown  in Fig.\,\ref{fig:modelData}.a (blue circles).
    As a result, our simulations are correct only for high $m_F$ states. 
    For example, for $m_F=+4$ the simulations predict that after applying additional field gradient by the auxiliary coil pair,  the initial velocity of atoms which can be efficiently decelerated rises by up to 20\,m/s. 
    With the increase of the initial velocity, less atoms from low magnetic sublevels experience sufficient deceleration before they are pumped to the $m_F=+4$ state which reduces the net process efficiency and causes significant deviation of the experimental results  from the model.

    \item The maximum number of trapped atoms for the 7-beam MOT configuration.
    According to the simulations, the  gain in the number of atoms after adding the  ``Zeeman'' beam  should be  much larger than observed in the experiment. 
    Besides the factor described in the previous paragraph, the  discrepancy can also be caused by improper model of the velocity distribution of atoms compared to the  actual atomic beam.
    
    \item The results for the  6-beam MOT  (Fig.\,\ref{fig:motNatoms}.a and Fig.\,\ref{fig:modelData}.b). 
    Here, we observe the mismatch of both the optimal magnetic field gradient and MOT beams frequency detuning.
    Besides the issue with the Zeeman levels structure, we attribute this disagreement to a nontrivial orientation of the quantization axis in the MOT, and hence the  spatial  variation  of the $\sigma^+$ polarization component intensity. 
\end{enumerate}

The processes described in S3 and S4 lead to losses of atoms from the cooling cycle since the laser radiation is only resonant with the $\ket{g,F=4}\rightarrow\ket{b,F=5}$ transition. 
However, the probabilities of these processes are  small \cite{Sukachev2010}, so  they do not impact the results significantly. 
In the experiment, we observe a 1.5 times increase of the atoms in MOT when adding a weak (0.2\,mW) $\ket{g,F=4,m_F=3}\rightarrow\ket{b,F=5,m_F=4}$ repump radiation. 
It  repumps  atomic population from the  $\ket{g,F=3}$ state. 
Our observations agree  with the effect of the repumping beam from Ref.~\cite{Sukachev2010}.

\end{document}